\documentclass[11pt,twoside]{article}

\usepackage{asp2006}
\usepackage{epsf}
\usepackage{lscape}

\begin{document}
\title{Near-infrared Polarimetry toward the Galactic Center\\
-- Magnetic Field Configuration in the Central One Degree Region --}   
\author{Shogo Nishiyama\altaffilmark{1}, Hirofumi Hatano\altaffilmark{2},
Motohide Tamura\altaffilmark{2}, and Tetsuya Nagata\altaffilmark{1}}   
\altaffiltext{1}{Department of Astronomy, Kyoto University, Kyoto 606-8502, Japan}
\altaffiltext{2}{Department of Astrophysics, Nagoya University, Nagoya, 464-8602, Japan}
\altaffiltext{3}{National Astronomical Observatory of Japan, Mitaka, Tokyo, 181-8588, Japan}

\begin{abstract} 
We present a NIR polarimetric map of the $1\deg \times 1\deg$ region toward the Galactic center.
Comparing Stokes parameters between highly reddened stars and less reddened ones, 
we have obtained a polarization originating from magnetically aligned dust grains 
at the central region of our Galaxy. The distribution of position angles shows a peak
at the parallel direction to the Galactic plane, suggesting a toroidal magnetic field configuration.
However, at high Galactic latitudes, the peak of the position angles departs from 
the direction of the Galactic plane. 
This may be a transition of a large-scale magnetic field configuration from toroidal to poloidal.
\end{abstract}

Polarization vectors of intrinsically un-polarized stars trace 
the plane-of-the-sky projection of the interstellar magnetic field (MF),
and polarimetric measurements of the stars of different distances can reveal
the three-dimensional distribution of the MF orientation.
To investigate the interstellar MF configuration at the Galactic center (GC),
we conducted NIR polarimetric observations of the central $1\deg \times 1\deg$ region
using NIR polarimeter SIRPOL on the IRSF telescope.
From the stars at the close side in the Galactic bulge, we can obtain polarization,
which is affected mainly by interstellar dust in the Galactic disk.
The light from stars at the far side in the bulge is transmitted through the dust
in the disk and the bulge. Using stars in both sides,
we can obtain the GC component of the interstellar polarization
(see Nishiyama et al. 2009 more detail).

In Fig. 1, we present the MF configuration at the GC derived by NIR polarimetry.
Our result is fairly consistent with those derived by previous sub-millimeter observations.
The histogram of the polarization angles ($\theta$) has a peak at the direction
almost parallel to the Galactic plane, suggesting a large-scale toroidal MF geometry 
in the Central Molecular Zone.

We find a systematic transition of the MF direction along the Galactic latitude.
Although no systematic change in $\theta$ is seen along the Galactic longitude,
clear change is seen along the Galactic latitude.
Near the Galactic plane ($|b|<0\fdg4$), the MF directions are almost parallel to the Galactic plane.
At higher Galactic latitudes, however, the peak of $\theta$ departs from the direction of the Galactic plane. 
This systematic change may indicate the transition of the large-scale MF configuration
in the central region of our Galaxy, at $|b| \sim 0\fdg4$.

\acknowledgements 

SN is financially supported by 
the Japan Society for the Promotion of Science (JSPS) 
through the JSPS Research Fellowship for Young Scientists.

\begin{center}
  \begin{figure}[h]
    \plotone{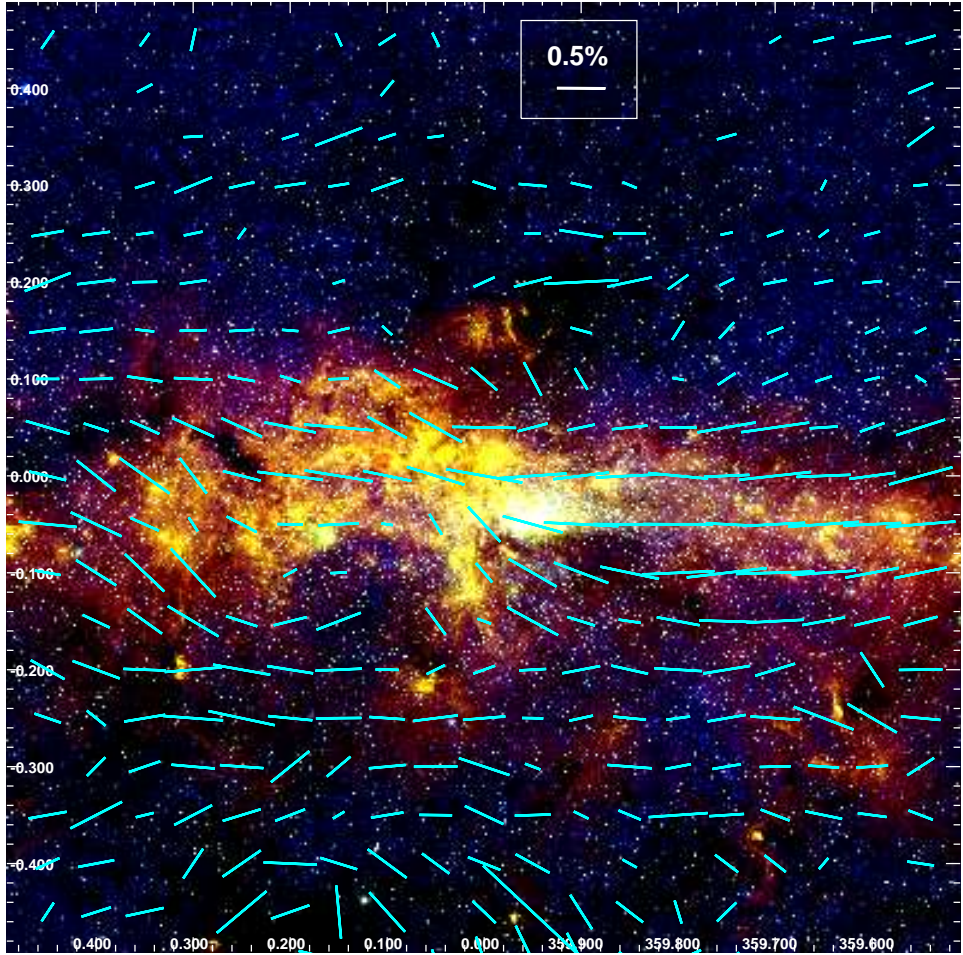}
    \caption{
      The magnetic field configuration at the Galactic center ($1\deg \times 1\deg$)
      derived by $K_S$-band polarization measurements (cyan bars) 
      shown together with an infrared combined image of 
      IRSF/SIRIUS in the $K_S$ band (blue), 
      and SST/IRAC in 5.8 $\mu$m (green) and 8.0 $\mu$m (red).
      Each bar is drawn parallel to the inferred magnetic field direction,
      and the length of the bar indicates the measured degree of polarization.
      Only vectors with $P/\delta P \geq 2$ are plotted.
    }
  \end{figure}
\end{center}

\end{document}